\documentclass[aps,prd,reprint,amsfonts,amssymb,amsmath,showpacs,preprintnumbers, letterpaper,nofootinbib,urlcolor=black,linkcolor=black]{revtex4-1}

\usepackage{array}
\usepackage[normalem]{ulem}

\usepackage{cancel}
\usepackage{braket,bm}
\usepackage[pass]{geometry} 
\usepackage{color}
\usepackage{mathrsfs,mathtools}
\usepackage{graphicx}
\usepackage{bbm}
\usepackage{enumerate}
  \DeclareMathAlphabet{\mathpzc}{OT1}{pzc}{m}{it}
\usepackage{subfigure}
\usepackage{graphicx}  

\usepackage[colorlinks,citecolor=black]{hyperref}

\def\d{\mathrm{d}}
\def\e{\mathrm{e}}

\def\vec{\mathbf}

\begin{document}

\title{A note on the novel 4D Einstein-Gauss-Bonnet gravity}
\author{Wen-Yuan Ai}
\email{wenyuan.ai@uclouvain.be}

\affiliation{Centre for Cosmology, Particle Physics and Phenomenology,\\ Université catholique de Louvain, Louvain-la-Neuve B-1348, Belgium}

\preprint{CP3-20-16}

\begin{abstract}
Recently, a novel 4D Einstein-Gauss-Bonnet gravity has been proposed by Glavan and Lin [Phys. Rev. Lett. 124, 081301 (2020)] by rescaling the coupling $\alpha \rightarrow \alpha/(D-4)$ and taking the limit $D\rightarrow 4$ at the level of equations of motion. This prescription, though was shown to bring non-trivial effects for some spacetimes with particular symmetries, remains mysterious and calls for scrutiny. Indeed, there is no continuous way to take the limit $D\rightarrow 4$ in the higher $D$-dimensional equations of motion because the tensor indices depend on the spacetime dimension and behave discretely. On the other hand, if one works with four-dimensional spacetime indices the contribution corresponding to the Gauss-Bonnet term vanishes identically in the equations of motion. A necessary condition (but may not be sufficient) for this procedure to work is that there is an embedding of the four-dimensional spacetime into the higher $D$-dimensional spacetime so that the equations in the latter can be properly interpreted after taking the limit. In this note, working with 2D Einstein gravity, we show several subtleties when applying the method used in [Phys. Rev. Lett. 124, 081301 (2020)]. 
\end{abstract}


\maketitle

\section{Introduction}
\label{sec:intro}
Although General Relativity (GR) is the most established and successful theory of gravity, it must be modified~\cite{Nojiri:2010wj,Clifton:2011jh,Capozziello:2011et,Nojiri:2017ncd,Ishak:2018his}. This is partially because GR is not theoretically complete and partially because several experimental observations, which are closely related to gravitational interaction, cannot be explained by it. Quantum effects, as typically shown in String theory, are believed to generate higher-order curvature terms in the low-energy effective theory of gravity. The most general metric theory of gravity which yields conserved second-order equations of motion in an arbitrary $D$-dimensional spacetime is given by Lovelock theory~\cite{Lovelock:1971yv}. The Lagrangian of Lovelock theory is given by a sum of terms with each term, $\mathcal{L}_{(n)}$ ($2n<D$), being the (generalized) Euler density in $2n$-dimensional spacetime. For the critical dimension of spacetime $D=2n$, $\mathcal{L}_{(n)}$ becomes topological and does not contribute to local dynamics. For example, when $D=4$, the Gauss-Bonnet term, $\mathcal{L}_{(2)}$, has no local dynamics and Lovelock theory reduces to GR. For studies of gravity with the Gauss-Bonnet term, see e.g. Refs.~\cite{Boulware:1985wk,Wiltshire:1985us,Wheeler:1985nh,Cai:1998vy,Cai:2001dz,Cvetic:2001bk,Odintsov:2019clh,Odintsov:2020sqy,Odintsov:2020zkl}.

Very recently, Glavan and Lin~\cite{Glavan:2019inb} proposed a novel four-dimensional (4D) Einstein-Gauss-Bonnet gravity where the Gauss-Bonnet term does contribute to local dynamics. To extract the local dynamics, they first rescale the coupling associated with the Gauss-Bonnet term in $D$-dimensional spacetime, $\alpha\rightarrow \alpha/(D-4)$, and then take the limit $D=4$ in the equations of motion. Through this process, they were able to obtain finite contributions from the Gauss-Bonnet term in the local equations of motion for some four-dimensional spacetimes with particular symmetries. This prescription has also been used earlier in Refs.~\cite{Tomozawa:2011gp,Cognola:2013fva} and below we shall refer to it as the {\it dimensional-regularization prescription}. Although the proposed 4D Einstein-Gauss-Bonnet gravity has already intrigued a large amount of work in applications (see Refs.~\cite{Konoplya:2020bxa,Guo:2020zmf,Fernandes:2020rpa,Wei:2020ght,Konoplya:2020qqh,Hegde:2020xlv,Kumar:2020owy,Ghosh:2020vpc,Doneva:2020ped,Zhang:2020qew,Konoplya:2020ibi,Singh:2020xju,Ghosh:2020syx,Konoplya:2020juj,Kumar:2020uyz,Zhang:2020qam,HosseiniMansoori:2020yfj,Wei:2020poh,Singh:2020nwo,Churilova:2020aca,Islam:2020xmy,Kumar:2020xvu,Liu:2020vkh,Mishra:2020gce,Li:2020tlo,Konoplya:2020cbv,Heydari-Fard:2020sib,Jin:2020emq,Zhang:2020sjh,EslamPanah:2020hoj,NaveenaKumara:2020rmi,Aragon:2020qdc,Malafarina:2020pvl}), the dimensional-regularization prescription has not been justified with the matched rigor. In this note, we examine it for the simplest case, i.e. two-dimensional (2D) Einstein gravity, and show that, without care, one may give incorrect interpretations in the obtained equations of motion. In the next section, we review the dimensional-regularization prescription used in Ref.~\cite{Glavan:2019inb}. In Sec.~\ref{sec:2DEinstein}, we apply the dimensional-regularization prescription to 2D Einstein gravity. Section~\ref{sec:conclusions} is left for discussions and conclusions.

\section{Dimensional-regularization prescription}
\label{sec:DR}

It is well known that the Gauss-Bonnet term is topological in four-dimensional spacetime while it becomes local in higher-dimensional spacetime. The integral of the Gauss-Bonnet invariant over a four-dimensional spacetime $\mathcal{M}_4$ (properly compactified) gives the Euler characteristic $\chi(\mathcal{M}_4)$ via
\begin{align}
\chi (\mathcal{M}_4)=\frac{1}{32\pi^2}\int_{\mathcal{M}_4}\d^4 x\sqrt{-g}\,\mathcal{G},
\end{align}
where
\begin{align}
 \mathcal{G}=R^{\mu\nu}_{\ \ \rho\sigma} R^{\rho\sigma}_{\ \ \mu\nu}-4 R^{\mu\nu}R_{\mu\nu} + R^2 =6R^{\mu\nu}_{\ \ [\mu\nu}R^{\rho\sigma}_{\ \ \rho\sigma]}
\end{align} 
is the Gauss-Bonnet invariant. The action from a topological term is thus invariant under the variation of the metric field whose boundary values are fixed. Thus, the Gauss-Bonnet term does not contribute to local dynamics. Further, in classical gravity, the topology of the spacetime is fixed which makes the Gauss-Bonnet term totally unobservable. (While in quantum gravity, one may sum over geometries with different topologies which might lead to observable effects.) However, the Gauss-Bonnet term becomes local when going beyond four-dimensional spacetime and does contribute to local dynamics in the equations of motion. One may expect that all the contributions from the Gauss-Bonnet term in Einstein's equations in higher $D$-dimensional spacetime carry a factor of $D-4$ so that they vanish when $D=4$.

To extract finite contributions to the local dynamics from the Gauss-Bonnet term, the authors in Ref.~\cite{Glavan:2019inb} rescale the coupling $\alpha$ to $\alpha/(D-4)$ which leads to the following action 
\begin{align}
S_{\rm EGB}=\int\d^D x\sqrt{-g}\left[\frac{M_{\rm P}^2}{2}R-\Lambda+\frac{\alpha}{D-4}\mathcal{G}\right].
\end{align}
While one still keeps the general $D$ in deriving the equations of motion (obviously, it makes no sense to put $D=4$ in the action), one may be able to take the limit $D=4$ finally. It was shown explicitly that for some highly symmetric spacetimes, the factor $1/(D-4)$ from the new action will cancel out all the factors $D-4$ in Einstein equations, giving rise to finite contributions after taking $D=4$~\cite{Glavan:2019inb}. The cancellation was also expected for more general cases.

If the dimensional-regularization prescription does work in general, it would be astonishing because one can then apply the same trick to other Lovelock densities in Lovelock gravity in $D$-dimensional spacetime and obtain finite contributions to the local dynamics in lower-dimensional spacetime which would be otherwise vanishing without playing the trick~\cite{Casalino:2020kbt}. To see it, we recall the general Lovelock Lagrangian
\begin{align}
\label{eq:lovelock}
\mathcal{L}=\sum_{n=0}^t\mathcal{L}_{(n)}=\sqrt{-g}\sum_{n=0}^t \alpha_n\mathcal{R}_{(n)}
\end{align}
where 
\begin{align}
\mathcal{R}_{(n)}=\frac{1}{2^n}\delta^{\mu_1\nu_1...\mu_n\nu_n}_{\alpha_1\beta_1...\alpha_n\beta_n}\prod_{r=1}^n R^{\alpha_r\beta_r}_{\ \ \ \ \ \mu_r\nu_r},
\end{align}
when multiplied by $\sqrt{-g}$, are the generalized Euler densities in $2n$-dimensions (also called as the Lovelock densities). Here 
\begin{align}
\delta^{\mu_1\nu_1...\mu_n\nu_n}_{\alpha_1\beta_1...\alpha_n\beta_n}=n!\,\delta^{\mu_1}_{[\alpha_1}\delta^{\nu_1}_{\beta_1}...\delta^{\mu_n}_{\alpha_n}\delta^{\nu_n}_{\beta_n]}
\end{align}
is the generalized Kronecker delta symbol. For example, $\mathcal{R}_{(0)}=1$, $\mathcal{R}_{(1)}=R$, $\mathcal{R}_{(2)}=\mathcal{G}$, giving the cosmological constant, the Hilbert-Einstein and the Gauss-Bonnet terms, respectively. In Eq.~\eqref{eq:lovelock}, $t=D/2$ for even $D$ and $t=(D-1)/2$ for odd $D$. This is simply because there is no nonvanishing $p$-form for $p>D$ in $D$-dimensional spacetime (the generalized Kronecker delta vanish for $2n>D$).

Every term $\mathcal{L}_{(n)}$ has a corresponding geometrical interpretation as the Gauss-Bonnet invariant does in four-dimensional spacetime. In $2n$-dimensional compact spacetime $\mathcal{M}_{2n}$ we have
\begin{align}
\chi(\mathcal{M}_{2n})=\frac{1}{(4\pi)^n n!}\int_{\mathcal{M}_{2n}}\d^{2n}x\,\sqrt{-g}\, \mathcal{R}_{(n)},
\end{align} 
where $\chi(M_{2n})$ is the Euler characteristic. Thus, the term $\mathcal{L}_{(n)}$ is topological in $2n$-dimensional spacetime (and hence does not contribute to any local dynamics) while becomes local in $D>2n$ dimensional spacetime. Therefore one may also expect that the contributions in the equations of motion in $D>2n$ dimensional spacetime from $\mathcal{L}_{(n)}$ are proportional to $D-2n$. One can then absorb $D-2n$ into $\alpha_n$ to generate finite local dynamics from the topological term in $2n$-dimensional spacetime. In particular, the method may be applied for Einstein gravity, with which we are very familiar, in two-dimensional spacetime.

One can further extend this procedure to gauge theories when there are terms that are topological in certain dimensional spacetime but cease to be so in higher-dimensional spacetime. For instance, the second Chern-form ${\rm tr} F_{\mu\nu}\widetilde{F}^{\mu\nu}$, which plays an important role in  quantum chromodynamics (QCD), is topological in four-dimensional spacetime. 

Since the consequences of the dimensional-regularization prescription, i.e. generating local dynamics from topological terms, are so novel, it perhaps requires more rigorous scrutiny. The very crucial condition for the dimensional-regularization prescription to work is that the continuous limit $D\rightarrow 4$ can be properly taken in the higher-dimensional equations of motion. Although this is believed and assumed by Glavan and Lin, to our opinion, it has not been proved explicitly.\footnote{As noted in Ref.~\cite{Glavan:2019inb}, one support on this assumption is the Einstein-Lovelock equations in terms of differential forms~\cite{Mardones:1990qc}, presented in equation (5) in that paper. However, that differential-form equation can only be read correctly for $D>2p$, as indicated by the indices and by itself cannot be applied to the limit $D=2p$.} (Otherwise, one can immediately apply the dimensional-regularization prescription in 4D Einstein-Gauss-Bonnet gravity for any spacetime, without the help of particular symmetries.) In the next section, we examine the dimensional-regularization prescription in the simplest case, the 2D Einstein gravity.

\section{2D Einstein Gravity}
\label{sec:2DEinstein}

We consider 2D Einstein gravity
\begin{align}
S=\int \d^2 x \sqrt{-g} \left[\alpha R-\Lambda+\mathcal{L}_{\rm matter}\right].
\end{align}
It has been known that 2D Einstein gravity, unless additional non-minimal coupling to other fields is included, is trivial. This is because $\sqrt{-g}R$ is a total derivative  in two-dimensional spacetime, precisely as the case for the Gauss-Bonnet term in four-dimensional spacetime. Taking the variation with respect to the metric, one obtains
\begin{align}
\Lambda g_{\mu\nu}=T_{\mu\nu},
\end{align}     
where 
\begin{align}
T_{\mu\nu}=\frac{2}{\sqrt{-g}}\frac{\delta \left(\sqrt{-g}\mathcal{L}_{\rm matter}\right)}{\delta g^{\mu\nu}}.
\end{align}
Taking the trace, one has $\Lambda=T/2$ where $T$ is the trace of the energy-momentum tensor. For a two-dimensional conformal field, whose energy-momentum tensor is traceless, a cosmological term is thus inconsistent.  

Now we want to generate local dynamics from the Einstein-Hilbert term in two-dimensional spacetime, following Ref.~\cite{Glavan:2019inb}. Quite interestingly, the obtained theory is equivalent to what obtained much earlier by Mann and Ross~\cite{Mann:1991qp,Mann:1992ar}. Although they also suggested deriving the theory via rescaling the coupling and taking the $D\rightarrow 2$ limit for GR, their theory can be formulated as one in which there is a scalar field non-minimally coupled to gravity and no divergent coupling at the {\it action} level. Thus their theory is solidly based and does not crucially rely on the dimensional-regularization prescription.  More details can be found in Refs.~\cite{Mann:1991md,Ohta:1996wq,Mann:1996cb,Mureika:2012fq,Frassino:2015oca}. Here we will take this simple example to draw some caution on the application of the dimensional-regularization prescription where one takes the $D\rightarrow 2$ limit at the level of equations of motion.

We now consider the $D$-dimensional theory
\begin{align}
\label{eq:GLEinstein}
S=\int\d^D x\sqrt{-g}\left[\frac{\alpha}{D-2}R-\Lambda+\mathcal{L}_{\rm matter}\right].
\end{align}
Einstein field equations read
\begin{align}
\label{eq:DEinstein}
\frac{2\alpha}{D-2}\left(R_{\mu\nu}-\frac{1}{2}Rg_{\mu\nu}\right)+\Lambda g_{\mu\nu}=T_{\mu\nu}.
\end{align}
Now taking the limit $D\rightarrow 2$ from above,  i.e. from $D>2$, in the Einstein equation is subtle, if not ill-defined. Even the limit $D\rightarrow 2$ can be understood in a continuous sense for some explicit factors of $D-2$ appearing in the equations of motion, the indices behave discretely. For example, there are $D(D-1)/2$ independent components\footnote{$D$ of them are constraint equations.} in the Einstein equation and the metric tensor. How these equations continuously evolve from the higher $D$-dimensional case to the two-dimensional case in which we have only one independent metric component? {For instance, what does it mean by taking $D=2.1$ in Eq.~\eqref{eq:DEinstein} if we are concerned about the indices?}\footnote{In the dimensional regularization in quantum field theory, one also encounters quantities carrying indices, e.g. momenta. But there one typically takes analytic continuation of $D$ in the results of the loop integrals which carry no tensor index. For a rigorous treatment, see e.g. Ref.~\cite{Collins:1984xc}.} If we assume that the indices take values as in a two-dimensional spacetime before taking the limit, then the Einstein tensor vanishes identically. Anyway, the indices cannot take a continuous value. The problem pointed out here is general for the dimensional-regularization prescription, not only specific to 2D Einstein gravity. We refer to it as the {\it index problem}.   

{Whether or not one can extract a total factor $D-2$ in the Einstein tensor in $D$-dimensional spacetime is unimportant. For an integer $D\geq 3$, defining a tensor $H_{\mu\nu}$ through $G_{\mu\nu}\equiv (D-2)H_{\mu\nu}$ is trivial; $H_{\mu\nu}$ has exactly the same tensor structure as the Einstein tensor. Even if we ignore the index problem when we take the limit $D\rightarrow 2$, it is mysterious if $G_{\mu\nu}$ vanishes while $H_{\mu\nu}$ does not for $D=2$ when they have the same tensor structure.}

Then in which cases may the limit have physical interpretations? First, for scalar equations, e.g. the trace of the Einstein equation, there is not the index problem. For the general tensor equations, we have to first embed the two-dimensional spacetime into the $D$-dimensional spacetime so that there is a clear map between the components of the metric tensors and the Einstein equations in the higher- and lower-dimensional cases. {Consider a two-dimensional spacetime with coordinates $\{x^0,x^1\}$ embedded in a $D$-dimensional spacetime with coordinates $\{x^0,x^1,\vec{x}^a\}$ where $a=2,..., D-1$. Then one may take the $D\rightarrow 2$ limit for the zero-zero, zero-one and one-one components in the Einstein equation.}  After taking the limit, one may simply discard the equations for the components from the extra dimensions. {Note that, the dimensional-regularization prescription differs from the dimensional reduction through Kaluza-Klein compactification. In the former the extra dimensions have no physical meaning and only serve to define the limit~\cite{Glavan:2019inb}. } For the example spacetimes that have been worked out in Ref.~\cite{Glavan:2019inb} in which the metric is given through an Ansatz which is valid for both the higher- and lower-dimensional cases, the embedding is automatically assumed.   For example, when constructing a black hole solution in the 4D Einstein-Gauss-Bonnet gravity from the 5D solution, it was assumed that the three-space of the four-dimensional spacetime lies on the equatorial hyperplane in the four-space of the five-dimensional spacetime. Our question is, does the obtained solution through the dimensional-regularization prescription describe a four-dimensional black hole faithfully or an effective one viewed on the equatorial hyperplane in the five-dimensional Schwarzschild spacetime? The full answer will not be pursued in this note but we will show the possibility for the latter interpretation through 2D Einstein gravity.


Taking the trace of Eq.~\eqref{eq:DEinstein}, we obtain 
\begin{align}
\label{eq:DEinstein-trace}
-\alpha R+D\Lambda=T.
\end{align}
This equation is correct for $D>2$ but now the limit $D=2$ can be taken and gives
\begin{align}
\label{eq:2D-trace}
-\alpha R+2\Lambda=T
\end{align} 
without causing any trouble. Do we obtain a non-trivial local dynamics for 2D Einstein gravity? Yes and no. For the yes, it is because in this case one may give a certain interpretation for the obtained equation of motion, as we shall show how later. For the no, it is because when we trace back to the origin Eq.~\eqref{eq:2D-trace} is really inherited from the higher-dimensional dynamics. To see it more clearly, assume we live in a $D=3$ dimensional world where the Einstein-Hilbert term contributes to the equation of motion (but no propagating gravitational modes). (Three-dimensional spacetime may not be a good example because 3D Einstein gravity can be formulated as Chern-Simons theory and hence is also topological, see Ref.~\cite{Witten:1988hc}.) Then Eq.~\eqref{eq:DEinstein-trace} gives the 3D trace equation
\begin{align}
-\alpha R+3\Lambda^{(3)}=T.
\end{align} 
If we pretend to take the limit $(D=3)\rightarrow 2$, we are simply redefining the cosmological constant $\Lambda=3\Lambda^{(3)}/2$.

Now if we restrict $R(t,x,y)$, $T(t,x,y)$ (with $\{t,x,y\}$ being a chosen coordinate frame) on the two-dimensional subspace $\{t,x\}$ and interpret $\Lambda\equiv 3\Lambda^{(3)}/2$ as an (effective) cosmological constant in this two-dimensional spacetime $\{t,x\}$, we can interpret Eq.~\eqref{eq:2D-trace} as describing 2D local gravity dynamics from the Einstein-Hilbert term. However, this is an illusion because we are really working with the higher-dimensional dynamics but look at a sub-spacetime embedded in it. If one wanders on a wire and gets surprised by the local gravity he/she sees, he/she may explain all this by just taking a step off that wire.

{The trace equation~\eqref{eq:DEinstein-trace} describes the full dynamics for the maximally symmetric spacetime.} Taking the limit $D=2$ for the vacuum case, one obtains $R=2\Lambda/\alpha$. As we argued, this maximally symmetric two-dimensional spacetime might be interpreted as a time-like two-surface in higher $D$-dimensional spacetime. Different $D$ gives different relation between $\Lambda^{(D)}$, the cosmological constant in $D$-dimensional spacetime, and the effective $\Lambda$.  We can also work with the Friedmann-Lema\^{i}tre-Robertson-Walker (FLRW) universe and the Schwarzschild spacetime, precisely as what the authors of Ref.~\cite{Glavan:2019inb} did for the 4D Einstein-Gauss-Bonnet gravity. With Ansatzs for the metric, one can reduce Einstein equations to scalar equations. 

For a $D$-dimensional FLRW metric $\d s^2=-\d t^2+a^2(t)[(1/(1-k r^2))\d r^2+r^2\d \Omega_{D-2}^2]$ with $\d\Omega_{D-2}^2$ being the metric of the unit $(D-2)$-sphere, Einstein equations~\eqref{eq:DEinstein} lead to the Friedman equation
\begin{align}
\label{eq:FriedmannD}
\alpha (D-1)\left(H^2+\frac{k}{a^2}\right)=\rho+\Lambda,
\end{align}
and the continuity equation
\begin{align}
\label{eq:continuity}
\dot{\rho}+(D-1)(\rho+p)H=0,
\end{align}
where $\rho=T_{00}$, $p=T_{ii}$ and $H=\dot{a}/a$ with $\dot{}$ represents the derivative with respect to $t$. {The derivation of the above equations is tedious and the details are given in Appendix~\ref{app:2DFLRW}.} One indeed can take the limit $D=2$. However, in this case even the embedding picture should be interpreted with care because the equations for $D>2$ is not related to $D=2$ by simple redefinitions of parameters because there is no other parameter except for the dimension $D$ in the continuity equation.

For Schwarzschild spacetime, we consider the following metric
\begin{align}
\label{eq:SchwarzschildMetric}
\d s^2=-\e^{2\omega(r)}\d t^2+\e^{-2\omega(r)}\d r^2+r^2\d\Omega_{D-2}^2.
\end{align} 
The dynamic equations could be derived similarly as in Appendix~\ref{app:2DFLRW}. {For example, we consider the vacuum Einstein equation with vanishing cosmological constant, $G_{\mu\nu}$=0, in $D$-dimensional spacetime. With the metric Ansatz~\eqref{eq:SchwarzschildMetric}, the component equations $G_{00}=0$ and $G_{11}=0$ give the same equation while $G_{ii}=0$ with $i\geq 2$ give another equation~\cite{Tangherlini:1963bw}. We finally have\footnote{In Ref.~\cite{Tangherlini:1963bw}, the metric signature is $(+,-,-,\cdots,-)$. But the equation of motion for $\omega(r)$ shall not depend on this choice of the metric signature.}
\begin{subequations} 
\begin{align}
\label{19a}
&\frac{(D-2)\e^{2\omega}}{2}\left(\frac{2\omega'}{r}+\frac{D-3}{r^2}\right)-\frac{(D-2)(D-3)}{2r^2}=0,\\
\label{19b}
&\frac{\e^{2\omega}}{2}\left[2\omega''+2\omega^{\prime 2}+\frac{(D-4)(D-3)}{r^2}\right]-\frac{(D-3)(D-4)}{2r^2}=0,
\end{align}
\end{subequations}
where the prime denotes the derivative with respect to $r$.} We see that for the $t$-$t$ and $r$-$r$ components of the Einstein equation (Eq.~\eqref{19a}), the factors $D-2$ will be cancelled out by the factor $1/(D-2)$ in Eq~\eqref{eq:DEinstein}. However for the other component equations (Eq.~\eqref{19b}), the divergent factor $1/(D-2)$ in Eq.~\eqref{eq:DEinstein} will not be cancelled out. This may by itself again indicate an inconsistency of the dimensional-regularization prescription. However, as we mentioned above, one may also simply discard the the other component equations beyond the $\{t,r\}$ dimensions when taking the $D\rightarrow 2$ limit. One could also construct 2D black hole solutions~\cite{Nojiri:2020tph}.

We believe the above analysis also applies to the 4D Einstein-Gauss-Bonnet gravity and therefore more checks beyond the highly symmetric spacetimes are needed. We emphasize that to define a theory, one shall be able to fully determine the dynamical equations without symmetry constraints on the spacetime. Otherwise, the embedding picture may be preferred. Note that the Gauss-Bonnet term does play a role in four-dimensional anti-de Sitter (AdS) spacetime in the context of AdS/CFT duality~\cite{Maldacena:1997re}. Its topological nature, i.e. being a boundary term, in the holographic renormalization correctly leads to the standard thermodynamics for AdS black holes~\cite{Miskovic:2009bm}.

\section{Conclusions and Discussions}
\label{sec:conclusions}

The novel 4D Einstein-Gauss-Bonnet gravity,  recently proposed by Glavan and Lin, has intrigued great interests in the community of gravity. In this theory, the topological Gauss-Bonnet term was shown to have local dynamics. The way to extract the local dynamics, however, relies on an unusual action principle where they have a coupling $\alpha /(D-4)$ associated with the Gauss-Bonnet term in the action, and take the limit $D=4$ in the equations of motion. One then expects that the divergent factor $D-4$ will be canceled out by the factors $D-4$ in the equations of motion. In this note, we first argue that, if this dimensional-regularization prescription does work, then it can be applied as a general method to extract local dynamics from topological terms, even beyond gravity theories. Second, we point out the index problem for this procedure. Specifically, even though the factors $D-4$ can be taken to the zero limit continuously, the tensor indices take discrete values and it is unclear how the equations of motion in a general $D$-dimensional spacetime will converge to the four-dimensional dynamical equations. One condition we argue is that the four-dimensional spacetime must be able to be embedded into the higher $D$-dimensional spacetime. Third, working with 2D Einstein gravity, we show a different interpretation on the obtained equations of motion from the dimensional-regularization prescription.

{If one constructs the four-dimensional dynamics through dimensional reduction, there must be additional degrees of freedom introduced in the effective four-dimensional theory. Indeed the effective 4D Einstein-Gauss-Bonnet theory can be reformulated as a scalar-tensor theory~\cite{Kobayashi:2020wqy,Lu:2020iav} in which an extra scalar field appears. It was shown that the linear perturbation of the vacuum, however, contains only the gravity~\cite{Ma:2020ufk}. The asymptotic structure for such a scalar-tensor theory has been studied in Ref.~\cite{Lu:2020mjp}. For different possible remedies of the dimensional-regularization prescription, which also typically lead to extra degrees of freedom, see Refs.~\cite{Fernandes:2020nbq,Aoki:2020lig}. At last, we note that, after the appearance of this note, more and more authors have raised doubts on the original 4D Einstein-Gauss-Bonnet gravity obtained from the dimensional-regularization prescription, see Refs.~\cite{Gurses:2020ofy,Shu:2020cjw,Mahapatra:2020rds,Hennigar:2020lsl,Tian:2020nzb,Bonifacio:2020vbk,Arrechea:2020evj}.}\\

{\it Note added:} After this work has been finished independently, the author noticed that the observation on 2D Einstein gravity from the dimensional-regularization prescription has also been made by Nojiri and Odintsov~\cite{Nojiri:2020tph}, whose paper came out on arXiv only several days ago (03, April) before the present work was submitted.

\section*{Acknowledgments}

We are grateful to Dra\v{z}en Glavan and Carlos Tamarit for helpful discussions. We also thank Robert Mann, Rodrigo Olea and Julio Oliva for kind comments. WYA is supported by the Incoming Postdoc fellowship of UC Louvain.

\begin{appendix}
\renewcommand{\theequation}{\Alph{section}\arabic{equation}}
\setcounter{equation}{0}

\section{Dynamic equations in 2D FLRW universe from the dimensional-regularization prescription}

\label{app:2DFLRW}
In this appendix we derive the dynamic equations in the 2D FLRW universe using the dimensional-regularization prescription. As we shall see, the dimensional-regularization prescription indeed yields regular equations of motion in this highly symmetric spacetime. But this by no means proves the validity of the dimensional-regularization prescription to define a {\it theory}. And the problems and subtleties mentioned in the main text should appear immediately if we go beyond the highly symmetric spacetimes.

Our main task is to compute the Einstein tensor in $D$-dimensional FLRW universe. Let us first consider the simpler flat case, $k=0$, with the metric $\d s^2=-\d t^2+a^2(t)(\d x_1^2+...\d x_{D-1}^2)$. It is easy to show that the nonvanishing Christoffel symbols are
\begin{align}
\Gamma^0_{\ ij}=\delta_{ij}a\dot{a},\quad \Gamma^i_{\ 0j}=\delta^i_j\frac{\dot{a}}{a},
\end{align}
where both $\delta_{ij}$ and $\delta^i_j$ are the Kronecker symbol and $i,j=1,...,D-1$. Using the formula
\begin{align}
R_{\mu\nu}=\Gamma^\alpha_{\ \mu\nu,\alpha}-\Gamma^\alpha_{\ \mu\alpha,\nu}+\Gamma^\alpha_{\ \beta\alpha}\Gamma^\beta_{\ \mu\nu}-\Gamma^\alpha_{\ \beta\nu}\Gamma^\beta_{\ \mu\alpha},
\end{align}
one obtains the nonvanishing  components of the Ricci tensor
\begin{align}
\label{eq:A3}
R_{00}=-(D-1)\frac{\ddot{a}}{a},\quad R_{ii}=(D-2)\dot{a}^2+a\ddot{a}.
\end{align}
The Ricci scalar is then
\begin{align}
R=(D-1)\left[(D-2)\left(\frac{\dot{a}}{a}\right)^2+2\frac{\ddot{a}}{a}\right].
\end{align} 
Finally, one obtains the nonvanishing components of the Einstein tensor
\begin{subequations}
\begin{align}
G_{00}&=\frac{(D-1)(D-2)}{2}\left(\frac{\dot{a}}{a}\right)^2,\\
G_{ii}&=-\frac{(D-2)(D-3)}{2}\dot{a}^2-(D-2)a\ddot{a}.
\end{align}
\end{subequations}
Substituting the above expressions into Eq.~\eqref{eq:DEinstein} and recalling $T_{00}=\rho$, $T_{ii}=pg_{ii}$, we obtain 
\begin{subequations}
\begin{align}
\label{A7}
\alpha (D-1)H^2&=\rho+\Lambda,\\
\label{A8}
-\alpha(D-3)\left(\frac{\dot{a}}{a}\right)^2-2\alpha\frac{\ddot{a}}{a}&=p-\Lambda.
\end{align}
\end{subequations}
Now taking the limit $D\rightarrow 2$, we obtain the dynamic equations in 2D flat FLRW universe in Einstein gravity from the dimensional-regularization prescription. Note that from Eqs.~\eqref{A7},~\eqref{A8}, one can obtain the continuity equation~\eqref{eq:continuity}.

Next we consider the general case $\d s^2=-\d t^2+a^2(t)[\d r^2/(1-kr^2)+r^2\d\Omega^2_{D-2}]$ where we used the coordinates $x^\mu=\{t,r,\theta_1,...\theta_{D-2}\}$. In this case, all the nonvanishing Christoffel symbols are
\begin{subequations}
\begin{align}
&\Gamma^0_{\ ii}=\frac{\dot{a}}{a}g_{ii},\ \Gamma^i_{\ 0i}=\Gamma^i_{\ i0}=\frac{\dot{a}}{a},\\ 
&\Gamma^1_{\ 11}=\frac{kr}{1-kr^2},\ 
\Gamma^1_{\ 22}=-r(1-kr^2),\\
&\Gamma^1_{\ ii}=-r(1-kr^2)(\sin^2\theta_1\cdots\sin^2\theta_{i-2}),\  {\rm for}\ i\geq 3,\\
&\Gamma^i_{\ i1}=\Gamma^i_{\ 1i}=\frac{1}{r},\ {\rm for}\ i\geq 2, \\
&\Gamma^i_{\ ik}=\Gamma^i_{\ ki}=\cot\theta_{k-1},\ {\rm for}\ i>k\geq 2,\\
&\Gamma^i_{\ (i+1)(i+1)}=-\sin\theta_{i-1}\cos\theta_{i-1},\ {\rm for}\ i\geq 2,\\
&\Gamma^i_{\ (i+k)(i+k)}=-\sin\theta_{i-1}\cos\theta_{i-1}\notag\\
&\ \quad \quad\times(\sin^2\theta_i\cdots\sin^2\theta_{i+k-2}),\ {\rm for}\ i\geq 2, k\geq 2.
\end{align}
\end{subequations}
Now calculating the components of the Ricci tensor is straightforward but tedious. First, one can show that $R_{\mu\nu}=0$ if $\mu\neq \nu$. For $R_{00}$, we obtain the same result as in Eq.~\eqref{eq:A3}. Although $R_{ii}$ have different expressions with different $i$, one can show that $g^{ii}R_{ii}$ (here there is no summation over $i$) has the same expression for any $i$. The result is
\begin{align}
\label{eq:A17}
g^{ii}R_{ii}=(D-2)\frac{\dot{a}^2}{a^2}+\frac{\ddot{a}}{a}+(D-2)\frac{k}{a^2},
\end{align} 
where there is no summation over $i$. This leads to that all the $(ii)$-component equations of Eq.~\eqref{eq:DEinstein} are equivalent. From the expression of $R_{00}$ in Eq.~\eqref{eq:A3} and Eq.~\eqref{eq:A17}, one obtains the Ricci scalar
\begin{align}
R=(D-1)\left[(D-2)\frac{\dot{a}^2}{a^2}+2\frac{\ddot{a}}{a}+(D-2)\frac{k}{a^2}\right].
\end{align}
This finally leads to
\begin{subequations}
\begin{align}
G_{00}&=\frac{(D-1)(D-2)}{2}\left(H^2+\frac{k}{a^2}\right),\\
g^{ii}G_{ii}&=-\frac{(D-2)(D-3)}{2}\left(H^2+\frac{k}{a^2}\right)-(D-2)\frac{\ddot{a}}{a},
\end{align}
\end{subequations}
where in the second equation there is no summation over $i$. Substituting the above equations {into} Eq.~\eqref{eq:DEinstein}, one then obtains the dynamic equations~\eqref{eq:FriedmannD} and~\eqref{eq:continuity}. (Again, the continuity equation can be derived from the original two dynamic equations after some simple algebra.)

\end{appendix}


\begin{thebibliography}{99}

\bibitem{Nojiri:2010wj} 
  S.~Nojiri and S.~D.~Odintsov,
  ``Unified cosmic history in modified gravity: from F(R) theory to Lorentz non-invariant models,''
  Phys.\ Rept.\  {\bf 505}, 59 (2011)
  [arXiv:1011.0544 [gr-qc]].

\bibitem{Clifton:2011jh}
T.~Clifton, P.~G.~Ferreira, A.~Padilla and C.~Skordis,
``Modified Gravity and Cosmology,''
Phys.\ Rept.\  \textbf{513}, 1-189 (2012)
[arXiv:1106.2476 [astro-ph.CO]].

\bibitem{Capozziello:2011et}
S.~Capozziello and M.~De Laurentis,
``Extended Theories of Gravity,''
Phys.\ Rept.\  \textbf{509}, 167-321 (2011)
[arXiv:1108.6266 [gr-qc]].

\bibitem{Nojiri:2017ncd} 
  S.~Nojiri, S.~D.~Odintsov and V.~K.~Oikonomou,
  ``Modified Gravity Theories on a Nutshell: Inflation, Bounce and Late-time Evolution,''
  Phys.\ Rept.\  {\bf 692}, 1 (2017)
  [arXiv:1705.11098 [gr-qc]].
  
\bibitem{Ishak:2018his}
M.~Ishak,
``Testing General Relativity in Cosmology,''
Living Rev. Rel. \textbf{22}, no.1, 1 (2019)
[arXiv:1806.10122 [astro-ph.CO]].


\bibitem{Lovelock:1971yv}
D.~Lovelock,
``The Einstein tensor and its generalizations,''
J.\ Math.\ Phys.\  \textbf{12}, 498-501 (1971)


\bibitem{Boulware:1985wk}
D.~G.~Boulware and S.~Deser,
``String Generated Gravity Models,''
Phys. Rev. Lett. \textbf{55}, 2656 (1985)

\bibitem{Wiltshire:1985us}
D.~Wiltshire,
``Spherically Symmetric Solutions of Einstein-maxwell Theory With a {Gauss-Bonnet} Term,''
Phys. Lett. B \textbf{169}, 36-40 (1986)

\bibitem{Wheeler:1985nh}
J.~T.~Wheeler,
``Symmetric Solutions to the Gauss-Bonnet Extended Einstein Equations,''
Nucl. Phys. B \textbf{268}, 737-746 (1986)

\bibitem{Cai:1998vy}
R.~Cai and K.~Soh,
``Topological black holes in the dimensionally continued gravity,''
Phys. Rev. D \textbf{59}, 044013 (1999)
[arXiv:gr-qc/9808067 [gr-qc]].

\bibitem{Cai:2001dz}
R.~Cai,
``Gauss-Bonnet black holes in AdS spaces,''
Phys. Rev. D \textbf{65}, 084014 (2002)
[arXiv:hep-th/0109133 [hep-th]].

\bibitem{Cvetic:2001bk}
M.~Cvetic, S.~Nojiri and S.~D.~Odintsov,
``Black hole thermodynamics and negative entropy in de Sitter and anti-de Sitter Einstein-Gauss-Bonnet gravity,''
Nucl. Phys. B \textbf{628}, 295-330 (2002)
[arXiv:hep-th/0112045 [hep-th]].

\bibitem{Odintsov:2019clh} S.~Odintsov and
V.~Oikonomou, 
``Inflationary Phenomenology of Einstein Gauss-Bonnet
Gravity Compatible with GW170817,'' Phys. Lett. B \textbf{797} (2019), 134874 
[arXiv:1908.07555 [gr-qc]].

\bibitem{Odintsov:2020sqy} S.~Odintsov,
V.~Oikonomou and F.~Fronimos, ``Rectifying Einstein-Gauss-Bonnet
Inflation in View of GW170817,'' [arXiv:2003.13724 [gr-qc]].

\bibitem{Odintsov:2020zkl}
S.~D.~Odintsov and V.~K.~Oikonomou,
``Swampland Implications of GW170817-compatible Einstein-Gauss-Bonnet Gravity,''
Phys. Lett. B \textbf{805}, 135437 (2020)
[arXiv:2004.00479 [gr-qc]].
 

\bibitem{Glavan:2019inb}
D.~Glavan and C.~Lin,
``Einstein-Gauss-Bonnet gravity in 4-dimensional space-time,''
Phys.\ Rev.\ Lett.\  \textbf{124}, no.8, 081301 (2020)
[arXiv:1905.03601 [gr-qc]].


\bibitem{Tomozawa:2011gp}
Y.~Tomozawa,
``Quantum corrections to gravity,''
[arXiv:1107.1424 [gr-qc]].

\bibitem{Cognola:2013fva}
G.~Cognola, R.~Myrzakulov, L.~Sebastiani and S.~Zerbini,
``Einstein gravity with Gauss-Bonnet entropic corrections,''
Phys.\ Rev.\ D \textbf{88}, no.2, 024006 (2013)
[arXiv:1304.1878 [gr-qc]].


\bibitem{Konoplya:2020bxa}
R.~Konoplya and A.~Zinhailo,
``Quasinormal modes, stability and shadows of a black hole in the novel 4D Einstein-Gauss-Bonnet gravity,''
[arXiv:2003.01188 [gr-qc]].

\bibitem{Guo:2020zmf}
M.~Guo and P.~C.~Li,
``Innermost stable circular orbit and shadow of the $4D$ Einstein–Gauss–Bonnet black hole,''
Eur. Phys. J. C \textbf{80}, no.6, 588 (2020)
[arXiv:2003.02523 [gr-qc]].


\bibitem{Fernandes:2020rpa}
P.~G.~S.~Fernandes,
``Charged Black Holes in AdS Spaces in $4D$ Einstein Gauss-Bonnet Gravity,''
Phys. Lett. B \textbf{805}, 135468 (2020)
[arXiv:2003.05491 [gr-qc]].



\bibitem{Wei:2020ght}
S.~Wei and Y.~Liu,
``Testing the nature of Gauss-Bonnet gravity by four-dimensional rotating black hole shadow,''
[arXiv:2003.07769 [gr-qc]].

\bibitem{Konoplya:2020qqh}
R.~A.~Konoplya and A.~Zhidenko,
``Black holes in the four-dimensional Einstein-Lovelock gravity,''
Phys. Rev. D \textbf{101}, no.8, 084038 (2020)
[arXiv:2003.07788 [gr-qc]].


\bibitem{Hegde:2020xlv}
K.~Hegde, A.~Naveena Kumara, C.~A.~Rizwan, A.~K.~M. and M.~S.~Ali,
``Thermodynamics, Phase Transition and Joule Thomson Expansion of novel 4-D Gauss Bonnet AdS Black Hole,''
[arXiv:2003.08778 [gr-qc]].

\bibitem{Kumar:2020owy}
R.~Kumar and S.~G.~Ghosh,
``Rotating black holes in the novel $4D$ Einstein-Gauss-Bonnet gravity,''
[arXiv:2003.08927 [gr-qc]].

\bibitem{Ghosh:2020vpc}
S.~G.~Ghosh and S.~D.~Maharaj,
``Radiating black holes in the novel 4D Einstein-Gauss-Bonnet gravity,''
[arXiv:2003.09841 [gr-qc]].

\bibitem{Doneva:2020ped}
D.~D.~Doneva and S.~S.~Yazadjiev,
``Relativistic stars in 4D Einstein-Gauss-Bonnet gravity,''
[arXiv:2003.10284 [gr-qc]].

\bibitem{Zhang:2020qew}
Y.~Zhang, S.~Wei and Y.~Liu,
``Spinning test particle in four-dimensional Einstein-Gauss-Bonnet Black Hole,''
[arXiv:2003.10960 [gr-qc]].

\bibitem{Konoplya:2020ibi}
R.~Konoplya and A.~Zhidenko,
``BTZ black holes with higher curvature corrections in the 3D Einstein-Lovelock theory,''
[arXiv:2003.12171 [gr-qc]].

\bibitem{Singh:2020xju}
D.~V.~Singh and S.~Siwach,
``Thermodynamics and P-v criticality of Bardeen-AdS Black Hole in 4-D Einstein-Gauss-Bonnet Gravity,''
[arXiv:2003.11754 [gr-qc]].

\bibitem{Ghosh:2020syx}
S.~G.~Ghosh and R.~Kumar,
``Generating black holes in the novel $4D$ Einstein-Gauss-Bonnet gravity,''
[arXiv:2003.12291 [gr-qc]].

\bibitem{Konoplya:2020juj}
R.~Konoplya and A.~Zhidenko,
``(In)stability of black holes in the 4D Einstein-Gauss-Bonnet and Einstein-Lovelock gravities,''
[arXiv:2003.12492 [gr-qc]].


\bibitem{Kumar:2020uyz}
A.~Kumar and R.~Kumar,
``Bardeen black holes in the novel $4D$ Einstein-Gauss-Bonnet gravity,''
[arXiv:2003.13104 [gr-qc]].

\bibitem{Zhang:2020qam}
C.~Zhang, P.~Li and M.~Guo,
``Greybody factor and power spectra of the Hawking radiation in the novel $4D$ Einstein-Gauss-Bonnet de-Sitter gravity,''
[arXiv:2003.13068 [hep-th]].


\bibitem{HosseiniMansoori:2020yfj}
S.~A.~Hosseini Mansoori,
``Thermodynamic geometry of novel 4-D Gauss Bonnet AdS Black Hole,''
[arXiv:2003.13382 [gr-qc]].

\bibitem{Wei:2020poh}
S.~W.~Wei and Y.~X.~Liu,
``Extended thermodynamics and microstructures of four-dimensional charged Gauss-Bonnet black hole in AdS space,''
Phys. Rev. D \textbf{101}, no.10, 104018 (2020)
[arXiv:2003.14275 [gr-qc]].

\bibitem{Singh:2020nwo}
D.~V.~Singh, S.~G.~Ghosh and S.~D.~Maharaj,
``Clouds of string in $4D$ novel Einstein-Gauss-Bonnet black holes,''
[arXiv:2003.14136 [gr-qc]].

\bibitem{Churilova:2020aca}
M.~Churilova,
``Quasinormal modes of the Dirac field in the novel 4D Einstein-Gauss-Bonnet gravity,''
[arXiv:2004.00513 [gr-qc]].

\bibitem{Islam:2020xmy}
S.~U.~Islam, R.~Kumar and S.~G.~Ghosh,
``Gravitational lensing by black holes in $4D$ Einstein-Gauss-Bonnet gravity,''
[arXiv:2004.01038 [gr-qc]].

\bibitem{Kumar:2020xvu}
A.~Kumar and S.~G.~Ghosh,
``Hayward black holes in the novel $4D$ Einstein-Gauss-Bonnet gravity,''
[arXiv:2004.01131 [gr-qc]].

\bibitem{Liu:2020vkh}
C.~Liu, T.~Zhu and Q.~Wu,
``Thin Accretion Disk around a four-dimensional Einstein-Gauss-Bonnet Black Hole,''
[arXiv:2004.01662 [gr-qc]].

\bibitem{Mishra:2020gce}
A.~K.~Mishra,
``Quasinormal modes and Strong Cosmic Censorship in the novel 4D Einstein-Gauss-Bonnet gravity,''
[arXiv:2004.01243 [gr-qc]].

\bibitem{Li:2020tlo}
S.~Li, P.~Wu and H.~Yu,
``Stability of the Einstein Static Universe in $4 D$ Gauss-Bonnet Gravity,''
[arXiv:2004.02080 [gr-qc]].

\bibitem{Konoplya:2020cbv}
R.~A.~Konoplya and A.~F.~Zinhailo,
``Grey-body factors and Hawking radiation of black holes in $4D$ Einstein-Gauss-Bonnet gravity,''
[arXiv:2004.02248 [gr-qc]].

\bibitem{Heydari-Fard:2020sib}
M.~Heydari-Fard, M.~Heydari-Fard and H.~Sepangi,
``Bending of light in novel 4$D$ Gauss-Bonnet-de Sitter black holes by Rindler-Ishak method,''
[arXiv:2004.02140 [gr-qc]].

\bibitem{Jin:2020emq}
X.~Jin, Y.~Gao and D.~Liu,
``Strong gravitational lensing of a 4D Einstein-Gauss-Bonnet black hole in homogeneous plasma,''
[arXiv:2004.02261 [gr-qc]].

\bibitem{Zhang:2020sjh}
C.~Zhang, S.~Zhang, P.~Li and M.~Guo,
``Superradiance and stability of the novel 4D charged Einstein-Gauss-Bonnet black hole,''
[arXiv:2004.03141 [gr-qc]].

\bibitem{EslamPanah:2020hoj}
B.~Eslam Panah and K.~Jafarzade,
``4D Einstein-Gauss-Bonnet AdS Black Holes as Heat Engine,''
[arXiv:2004.04058 [hep-th]].

\bibitem{NaveenaKumara:2020rmi}
A.~Naveena Kumara, C.~A.~Rizwan, K.~Hegde, M.~S.~Ali and A.~K.~M,
``Rotating 4D Gauss-Bonnet black hole as particle accelerator,''
[arXiv:2004.04521 [gr-qc]].

\bibitem{Aragon:2020qdc}
A.~Aragón, R.~Bécar, P.~González and Y.~Vásquez,
``Perturbative and nonperturbative quasinormal modes of 4D Einstein-Gauss-Bonnet black holes,''
[arXiv:2004.05632 [gr-qc]].

\bibitem{Malafarina:2020pvl}
D.~Malafarina, B.~Toshmatov and N.~Dadhich,
``Dust collapse in 4D Einstein-Gauss-Bonnet gravity,''
Phys. Dark Univ. \textbf{30}, 100598 (2020)
[arXiv:2004.07089 [gr-qc]].





\bibitem{Casalino:2020kbt}
A.~Casalino, A.~Colleaux, M.~Rinaldi and S.~Vicentini,
``Regularized Lovelock gravity,''
[arXiv:2003.07068 [gr-qc]].

\bibitem{Mardones:1990qc}
A.~Mardones and J.~Zanelli,
``Lovelock-Cartan theory of gravity,''
Class.\ Quant.\ Grav.\  \textbf{8}, 1545-1558 (1991)

\bibitem{Mann:1991qp}
R.~B.~Mann,
``Lower dimensional black holes,''
Gen.\ Rel.\ Grav.\  \textbf{24}, 433-449 (1992)

\bibitem{Mann:1992ar}
R.~B.~Mann and S.~Ross,
``The D $\rightarrow$ 2 limit of general relativity,''
Class.\ Quant.\ Grav.\  \textbf{10}, 1405-1408 (1993)
[arXiv:gr-qc/9208004 [gr-qc]].

\bibitem{Mann:1991md}
R.~B.~Mann, S.~Morsink, A.~Sikkema and T.~Steele,
``Semiclassical gravity in (1+1)-dimensions,''
Phys.\ Rev.\ D \textbf{43}, 3948-3957 (1991)

\bibitem{Ohta:1996wq}
T.~Ohta and R.~B.~Mann,
``Canonical reduction of two-dimensional gravity for particle dynamics,''
Class.\ Quant.\ Grav.\  \textbf{13}, 2585-2602 (1996)
[arXiv:gr-qc/9605004 [gr-qc]].

\bibitem{Mann:1996cb}
R.~B.~Mann and T.~Ohta,
``Exact solution for the metric and the motion of two bodies in (1+1)-dimensional gravity,''
Phys.\ Rev.\ D \textbf{55}, 4723-4747 (1997)
[arXiv:gr-qc/9611008 [gr-qc]].

\bibitem{Mureika:2012fq}
J.~Mureika and P.~Nicolini,
``Self-completeness and spontaneous dimensional reduction,''
Eur.\ Phys.\ J.\ Plus \textbf{128}, 78 (2013)
[arXiv:1206.4696 [hep-th]].

\bibitem{Frassino:2015oca}
A.~M.~Frassino, R.~B.~Mann and J.~R.~Mureika,
``Lower-Dimensional Black Hole Chemistry,''
Phys.\ Rev.\ D \textbf{92}, no.12, 124069 (2015)
[arXiv:1509.05481 [gr-qc]].

\bibitem{Collins:1984xc}
J.~C.~Collins,
``Renormalization,''
doi:10.1017/CBO9780511622656


\bibitem{Witten:1988hc}
E.~Witten,
``(2+1)-Dimensional Gravity as an Exactly Soluble System,''
Nucl.\ Phys.\ B \textbf{311}, 46 (1988)



\bibitem{Tangherlini:1963bw}
F.~Tangherlini,
``Schwarzschild field in n dimensions and the dimensionality of space problem,''
Nuovo Cim.\  \textbf{27}, 636-651 (1963)

\bibitem{Nojiri:2020tph}
S.~Nojiri and S.~D.~Odintsov,
``Novel cosmological and black hole solutions in Einstein and higher-derivative gravity in two dimensions,''
EPL \textbf{130}, no.1, 10004 (2020)
[arXiv:2004.01404 [hep-th]].


\bibitem{Maldacena:1997re}
J.~M.~Maldacena,
``The Large N limit of superconformal field theories and supergravity,''
Int. J. Theor. Phys. \textbf{38}, 1113-1133 (1999)
[arXiv:hep-th/9711200 [hep-th]].


\bibitem{Miskovic:2009bm}
O.~Miskovic and R.~Olea,
``Topological regularization and self-duality in four-dimensional anti-de Sitter gravity,''
Phys. Rev. D \textbf{79}, 124020 (2009)
[arXiv:0902.2082 [hep-th]].

\bibitem{Lu:2020iav}
H.~Lu and Y.~Pang,
``Horndeski Gravity as $D\rightarrow4$ Limit of Gauss-Bonnet,''
[arXiv:2003.11552 [gr-qc]].

\bibitem{Kobayashi:2020wqy}
T.~Kobayashi,
``Effective scalar-tensor description of regularized Lovelock gravity in four dimensions,''
JCAP \textbf{07}, 013 (2020)
[arXiv:2003.12771 [gr-qc]].


\bibitem{Ma:2020ufk}
L.~Ma and H.~Lu,
``Vacua and Exact Solutions in Lower-$D$ Limits of EGB,''
[arXiv:2004.14738 [gr-qc]].

\bibitem{Lu:2020mjp}
H.~Lu and P.~Mao,
``Asymptotic structure of Einstein-Gauss-Bonnet theory in lower dimensions,''
[arXiv:2004.14400 [hep-th]].

\bibitem{Fernandes:2020nbq}
P.~G.~S.~Fernandes, P.~Carrilho, T.~Clifton and D.~J.~Mulryne,
``Derivation of Regularized Field Equations for the Einstein-Gauss-Bonnet Theory in Four Dimensions,''
Phys. Rev. D \textbf{102}, no.2, 024025 (2020)
[arXiv:2004.08362 [gr-qc]].


\bibitem{Aoki:2020lig}
K.~Aoki, M.~A.~Gorji and S.~Mukohyama,
``A consistent theory of $D\rightarrow 4$ Einstein-Gauss-Bonnet gravity,''
[arXiv:2005.03859 [gr-qc]].

\bibitem{Gurses:2020ofy}
M.~Gurses, T.~C.~Sisman and B.~Tekin,
``Is there a novel Einstein-Gauss-Bonnet theory in four dimensions?,''
[arXiv:2004.03390 [gr-qc]].

\bibitem{Shu:2020cjw}
F.~W.~Shu,
``Vacua in novel 4D Einstein-Gauss-Bonnet Gravity: pathology and instability?,''
[arXiv:2004.09339 [gr-qc]].

\bibitem{Mahapatra:2020rds}
S.~Mahapatra,
``A note on the total action of $4D$ Gauss-Bonnet theory,''
[arXiv:2004.09214 [gr-qc]].

\bibitem{Hennigar:2020lsl}
R.~A.~Hennigar, D.~Kubiznak, R.~B.~Mann and C.~Pollack,
``On Taking the $D\to 4$ limit of Gauss-Bonnet Gravity: Theory and Solutions,''
JHEP {\bf 2007} (2020) 027
[arXiv:2004.09472 [gr-qc]].


\bibitem{Tian:2020nzb}
S.~Tian and Z.~H.~Zhu,
``Comment on "Einstein-Gauss-Bonnet Gravity in Four-Dimensional Spacetime",''
[arXiv:2004.09954 [gr-qc]].

\bibitem{Bonifacio:2020vbk}
J.~Bonifacio, K.~Hinterbichler and L.~A.~Johnson,
``Amplitudes and 4D Gauss-Bonnet Theory,''
Phys. Rev. D \textbf{102}, no.2, 024029 (2020)
[arXiv:2004.10716 [hep-th]].


\bibitem{Arrechea:2020evj}
J.~Arrechea, A.~Delhom and A.~Jiménez-Cano,
``Yet another comment on four-dimensional Einstein-Gauss-Bonnet gravity,''
[arXiv:2004.12998 [gr-qc]].

\end{thebibliography}
\end{document}